\documentclass[12pt]{article}
\usepackage{amsmath,amssymb} 
\usepackage{a4wide} 
\def\w{\langle W|}
\def\tildeW{\langle\widetilde{W} |}
\def\v{|V\rangle}
\def\tilde_V{|\widetilde{V}\rangle}

\def\bin_tilde_ED{\binom{\widetilde{E}}
                        {\widetilde{D}}}

\def\row_tilde_ED{\begin{pmatrix}
            \widetilde{E}\quad \widetilde{D}
           \end{pmatrix}}
\makeatletter
\@addtoreset{equation}{section}

\makeatother
\begin{document}
\title{Construction of a matrix product stationary state from solutions of finite size system}
\setcounter{footnote}{3}
\author{Y Hieida${}^{\dagger}$\footnote{e-mail:hieida@cc.saga-u.ac.jp}
\setcounter{footnote}{5}
 and T Sasamoto${}^{\ddagger}$\footnote{e-mail:sasamoto@stat.phys.titech.ac.jp}
\\
\\
${}^{\dagger}$
Computer and Network Centre, Saga University,\\
Saga 840-8502, Japan
\\
\\
${}^{\ddagger}$
Department of Physics, Tokyo Institute of Technology,\\
Oh-okayama 2-12-1, Meguro-ku, Tokyo 152-8551, Japan
}

\date{2004/03/09}
\maketitle

\begin{abstract}

Stationary states of stochastic models, which have $N$ states per site, in matrix product form are considered. First we
give a necessary condition for the existence of a finite $M$-dimensional matrix product state for any $\{N,M\}$. Second,
we give a method to construct the matrices from the stationary states of small size system when the above condition and
$N\le M$ are satisfied. Third, the method by which one can check that the obtained matrices are valid for any system
size is presented for the case where $M=N$ is satisfied.  The application of our methods is explained using three
examples: 
the asymmetric exclusion process, a model studied in [Jafarpour F H 2003 {\it J. Phys. A: Math. Gen.}{\bf 36} 7497] and
 a hybrid of both of the models.

\end{abstract}
\section{Introduction}

The one-dimensional many-body stochastic models have attracted much attention because of their rich nonequilibrium
behaviours and wide applicability in condensed matter physics, biology and other
fields.\cite{SZ-review,Privman-review,Hinrichsen-review,Schutz2003Review} 
The models also have fundamental aspects: they serve as models of which
we can inspect the behaviours and, through them, 
develop theories of nonequilibrium statistical physics.

To study the properties of such models, there are several methods.  One might start from doing Monte Carlo simulations
to get a rough grasp about what is going on. Then one might apply some approximate methods such as the mean field theory
or cluster approximation.  One might also adopt renormalization group techniques.  In many cases these kinds of analysis
give us satisfactory understanding of the models.

In some cases, however, these methods are not enough to fully understand the properties of the models. Monte Carlo
results and approximations 
sometimes disguise the physics  of the models.
For instance in \cite{RSS} it was shown that a numerically
apparent phase transition associated with a particle condensation is not accompanied by a non-analyticity of physical
quantities.  In such cases, it would be desirable to have exact solutions of the model. Of course most models do not
admit exact analytical treatment. But for some models, particularly for the ones in one dimension, one can obtain exact
solutions and study the properties of the models in quite detail.

In fact there is a class of models for which the exact stationary state can be written in a matrix product
form\cite{review-of-matrix-product-solution}. We abbreviate this matrix-product stationary state as MPSS in the
following.  After the discovery of an exact matrix-product groundstate for a one-dimensional quantum spin chain
\cite{AKLT1}, the first exact MPSS of the one-dimensional stochastic models was found in \cite{DEHP} for the asymmetric
simple exclusion process(ASEP) (\S\ref{sec:example-ASEP}).  It is interesting that numerical solutions in the matrix
product form (MPF) also play an important role
\cite{OR,RO}
in the method of density matrix renormalization group (DMRG)\cite{DMR}
and its higher-dimensional extension\cite{NRA1,NRA2} through a generalisation of MPF(tensor product form).

After \cite{DEHP}, the MPSSs have been found for a lot of
one-dimensional stochastic models.  
In addition, it is known that stationary states of stochastic Hamiltonians 
\begin{itemize}
 \item with nearest-neighbour interaction(this is the interaction in the
       current paper, treated at (\ref{eq:031210-1852})-(\ref{eq:031210-1855}) ) or
 \item with arbitrary, but finite, interaction range
\end{itemize}
can be expressed as {\it infinite} dimensional MPSSs(see \cite{Krebs-Sandow}
and \cite{Klauck-Schadschneider}, respectively).
So far, however, it has not been known how to construct systematically the
{\it finite } dimensional
MPSS for a model at hand.  One might try to find even a two
dimensional matrix product state very hard for several days in vain.  Hence, it would be desirable to have some methods
to know
\begin{enumerate}
 \item whether the stationary state of the model at hand can be written in a MPF or not,

 \item if yes, how to find the matrices in the MPF.
\end{enumerate}

The objective of this article is to partially answer this question.  More precisely, for the case where 
the dimensions 
($M$)
of the matrices are finite, 
we give
\begin{enumerate}
 \item for any $\{N,M\}$, a way to find
necessary conditions for 
the
existence of
an exact $M$-dimensional MPSS of stochastic models which have $N$ states per site

 \item a systematic way by which the exact concrete $M$-dimensional
       representation of the MPF can be constructed from the stationary
       states of small size system 
       if the condition obtained in the abovementioned step 1 is also a sufficient condition
       and $N\le M$ is satisfied

 \item for the case $M=N$, the method by which one can check that the obtained matrices are valid for any system size.
\end{enumerate}

This paper is organised as follows: In \S\ref{sec:method}, we describe our method.  Three examples are given in
\S\ref{sec:ex}.  The last section(\S\ref{sec:summary}) is devoted to the summary and a list of possible extensions of
our methods.

Hereafter, we omit the word ``exact'', namely, we use 
           ``a        MPSS'' and  ``a        representation''
instead of ``an exact MPSS'' and ``an exact representation'', respectively.

\section{Method}
\label{sec:method}

Let us consider a many-particle stochastic model on a chain of size $L$. Each site $j$ ($j=1,2,\cdots, L$) is assumed to
be in one of $N$ states which we denote as $\tau_j(=0,1,2,\dots, (N-1))$.  For instance when $N=2$, $\tau_j=0$
(resp. $\tau_j=1$) may represent a state in which the site $j$ is empty (resp. occupied by a particle). In this article
we only consider a model in continuous time setting, for which the dynamics can be described by the master equation,
\begin{equation}
\label{mas_eq}
 \frac{d}{dt} \vec{P}_L(t) = -H \vec{P}_L(t).
\end{equation}
Here $\vec{P}_L(t)$ is a vector whose component represents the probability of a system in a certain state at time $t$.
$H$ is a transition rate matrix describing the stochastic dynamics of the model.  This formulation of a stochastic model
is called a quantum Hamiltonian formalism of the master equation because of the obvious similarity of 
(\ref{mas_eq})
to the imaginary-time Schr\"odinger equation
\cite{SS}.
In a stationary state, to which we restrict our attention in the following, the left hand side of 
(\ref{mas_eq})
vanishes. The stationary state vector, $\vec{P}_L$, of the model(size $L$) is determined by 
\begin{equation}
 H \vec{P}_L = 0 .
 \label{eq:030626-0443}
\end{equation}

In this article, we are interested in a very special situation in which $\vec{P}_L$ can be written in a matrix product form (MPF),
\begin{equation}
 \vec{P}_L = \vec{P}^\mathrm{MPF}_L \quad\text{for}\quad L=1,2,3,\dots , 
\label{eq:030828-0227}
\end{equation}
where $\vec{P}^\mathrm{MPF}_L$ is the vector whose component $P^\mathrm{MPF}_L(\tau_1,\tau_2,\dots,\tau_L)$ is defined by
\begin{equation}
P^\mathrm{MPF}_L(\tau_1,\tau_2,\dots,\tau_L)
:=
\frac{1}{Z_L}\w A(\tau_1)A(\tau_2)\dots A(\tau_L)\v .
\label{eq:030828-0005}
\end{equation}
Here $\{A(\tau)\}_{\tau=0,1,\dots , (N-1)}$ are $M$-dimensional square matrices with $M$ being assumed to be
finite. $\w$ (resp. $\v$) is an  $M$-dimensional row(resp. column) vector. $Z_L$ is the normalisation constant defined by 
\begin{equation}
 Z_L := \w [A(0)+A(1)+\dots A(N-1)]^L\v . 
\label{eq:030828-0006}
\end{equation}
We will refer to $\{A(0),\cdots,A(N-1),\w,\v \}$ as ``a set of matrices'' for simplicity in the sequel. One should
notice that there is a trivial freedom of a similarity transformation for the choice of the matrices in
(\ref{eq:030828-0005}). 
If one introduces another set of matrices 
$\{\widetilde{A}(0),\cdots,\widetilde{A}(N-1),\tildeW,\tilde_V \}$ by
\begin{equation}
\begin{split}
 \langle W|S      &=: \langle \widetilde{W}|, \\
 S^{-1} A(\tau) S &=: \widetilde{A}(\tau), \quad (\tau=0,1,\cdots,N-1)\\
 S^{-1} |V\rangle &=: |\widetilde{V} \rangle ,
\end{split}
\label{eq:040218-1641}
\end{equation}
one has 
\begin{equation}
P^\mathrm{MPF}_L(\tau_1,\tau_2,\dots,\tau_L)
=
\frac{1}{Z_L}\tildeW \widetilde{A}(\tau_1)\widetilde{A}(\tau_2)
\dots \widetilde{A}(\tau_L)\tilde_V .
\label{eq:040218-1642}
\end{equation}

It is known that, $\vec{P}_L$ of some models has the MPF when some
  conditions of the model parameters are satisfied.

In the following discussions, we sometimes explain the main ideas using the $N=2$ case, in which case two matrices
$A(0),A(1)$ are renamed as $A(0)=: E,~ A(1)=: D$. Then the MPF may be written as 
\begin{equation}
 \vec{P}_L 
 = \frac{1}{Z_L}\w \binom{E}{D}^{\otimes L}\v 
 = \frac{1}{Z_L}\w \underbrace{\binom{E}{D}\otimes\binom{E}{D}\otimes \dots \binom{E}{D} }_{L \quad\text{times}} \v .
\end{equation}
For later use, we also define 
\begin{equation}
 P^{m,n} 
:=
 \frac{1}{Z_{m+n}} \langle W| \binom{E}{D}^{\otimes m}
 \left(E~ D\right)^{\otimes n} |V\rangle,
\label{eq:031212-0957}
\end{equation}
which is a conversion of the vector $\vec{P}_L$ to a matrix form with $L=m+n$. Notice that the rank of this matrix is at
most $2^{\text{min}\{m,n\}}$, where $\text{min}\{m,n\}$ is a function whose value is the smaller number of $m$ and $n$.

\subsection{How to find 
necessary conditions for 
the 
existence of an MPSS
}
\label{sec:method-how-to-find-the-condition}

In this subsection, we give a way to find 
necessary conditions 
for 
the existence of an $M$-dimensional MPSS.
The key observation here is
that
when the $\vec{P}_L$ has an $M$-dimensional 
MPF,
the rank of a matrix $P^{m,n}$ obtained from $\vec{P}_L$ is no larger than $M$. Hence, if we take $m$ and $n$ satisfying 
$N^m > M, N^n > M$, the rank of the matrix $P^{m, n}$ is $M$ which is strictly smaller than $N^{\text{min}\{m,n\}}$. We
call this phenomenon the ``rank deficiency''.  The rank deficiency occurs 
when the $\vec{P}_L$ has an $M$-dimensional 
MPF.
In reverse, it is by this rank deficiency that we can find 
necessary conditions
for the existence of an $M$-dimensional MPSS.

First we explain this using the case where $N=M=2,L=4$. Let us consider equation(\ref{eq:030626-0443}) for $L=4$ and 
suppose that the solution $\vec{P}_4$ has the MPF,
\begin{equation}
 \vec{P}_4 = \frac{1}{Z_4} \w \binom{E}{D}^{\otimes 4}\v.
\label{eq:030826-1900}
\end{equation}
We convert the vector form (\ref{eq:030826-1900}) into the matrix form,
\begin{equation}
P^{2,2} = \frac{1}{Z_4} \w \binom{E}{D}^{\otimes 2} \left (E\, D \right )^{\otimes 2}\v  ,
\label{eq:030828-1317}
\end{equation}
which is a 4$\times$4 matrix.

Now we show that, when $M=2$, the rank of $P^{2,2}$ is at most two, i.e., the rank deficiency of $P^{2,2}$ occurs. Since 
$\{EE|V\rangle, ED|V\rangle, DE|V\rangle, DD|V\rangle  \}$ is a set of $M(=2)$-dimensional vectors,
we can prepare two basis (column) vectors, which we call $|e_1\rangle, |e_2\rangle$.  There exist some coefficients,
$\{a_k,b_k\}_{k=1,2,3,4}$, such that
\begin{align}
 E E|V\rangle &= a_1 |e_1\rangle +b_1 |e_2\rangle ,\label{eq:021119-0101} \\
 E D|V\rangle &= a_2 |e_1\rangle +b_2 |e_2\rangle , \\
 D E|V\rangle &= a_3 |e_1\rangle +b_3 |e_2\rangle , \\
 D D|V\rangle &= a_4 |e_1\rangle +b_4 |e_2\rangle .
 \label{eq:021119-0100}
\end{align}
Using these, column vectors in the RHS of (\ref{eq:030828-1317}) are 
\begin{align}
         \langle W|\binom{E}{D}^{\otimes 2} EE |V\rangle
& =  a_1 \langle W|\binom{E}{D}^{\otimes 2}    |e_1\rangle
   + b_1 \langle W|\binom{E}{D}^{\otimes 2}    |e_2\rangle ,
\label{eq:021119-0112}\\
         \langle W|\binom{E}{D}^{\otimes 2} E D |V\rangle 
& =  a_2 \langle W|\binom{E}{D}^{\otimes 2} |e_1\rangle 
   + b_2 \langle W|\binom{E}{D}^{\otimes 2} |e_2\rangle ,
\label{eq:021119-0111}\\
         \langle W|\binom{E}{D}^{\otimes 2} D E |V\rangle 
& =  a_3 \langle W|\binom{E}{D}^{\otimes 2} |e_1\rangle 
   + b_3 \langle W|\binom{E}{D}^{\otimes 2} |e_2\rangle ,
\label{eq:021119-0110}\\
         \langle W|\binom{E}{D}^{\otimes 2} D D |V\rangle 
& =  a_4 \langle W|\binom{E}{D}^{\otimes 2} |e_1\rangle 
   + b_4 \langle W|\binom{E}{D}^{\otimes 2} |e_2\rangle .
\label{eq:021119-0109}
\end{align}
One notices that the right hand sides of (\ref{eq:021119-0112})-(\ref{eq:021119-0109}) are written as linear
combinations of the two vectors, 
\begin{equation}
\left\{ \langle W| \binom{E}{D}^{\otimes 2}|e_i\rangle \right\}_{i=1,2}.
\end{equation}
Hence there are at most two independent vectors among the four column vectors of $P^{2,2}$.  This means that the rank of
$P^{2,2}$ is at most two. The generalisation of the above argument to general $M,N,L$ case is not difficult.

One should notice that this can be used for checking the existence of a finite dimensional MPSS for a given model.  For
instance to be sure that there is no $(M=)$2 dimensional MPSS for a given parameter set of an $N=2$ model, all one needs
to do is to find a stationary state for $L=4$ and see that the rank deficiency does not occur.

Furthermore we can use this rank deficiency to find
necessary conditions
for 
the existence of an $M$-dimensional MPSS.
Algorithmically one performs the following steps.
\begin{enumerate}
 \item Solve equation(\ref{eq:030626-0443}) for $L$ for which one can take $m,n$ such that $L=m+n$ and $N^m > M, N^n > M$.
       
 \item Make the $N^m \times N^n$ matrix $P^{m,n}$ from $\vec{P}_{L}$.

 \item Calculate the rank of the matrix $P^{m,n}$.

 \item Find 
necessary conditions
for
the existence of an $M$-dimensional MPSS,
from conditions that the rank of $P^{m,n}$ is $M$.
\end{enumerate}
For instance, for the case where $M=N=2,L=4$, one finds $P_4$, converts it to $P^{2,2}$, computes the rank of it with
the model parameters unfixed and looks for a condition where the rank deficiency occurs. If one finds a condition where
the rank of $P^{2,2}$ is 2, it is a 
necessary condition
for
the existence of an $M(=2)$-dimensional MPSS.

Note that a necessary condition found in the step 4 may not be
a sufficient condition for the existence of a finite dimensional
MPSS.
In fact there occurs a different type of rank
deficiency, caused by the existence of algebraic
relations.  This really happens for the ASEP and will be explained at the end of \S\ref{sec:candidates-of-ASEP}.

In practice it is in general not an easy task to calculate the rank of a matrix with a lot of parameters unfixed. But if
we use a computer algebra system like Maple
(we have used Maple to perform most of the calculations in \S\ref{sec:ex}),
it is possible to do this to some extent. There would be several methods to 
compute the rank of a matrix on a computer. 
When explaining our methods in the next section,
we will adopt a version of 
Gaussian elimination procedure 
which consists of several elementary transformations.
By using elementary transformations, which do not change the rank of a 
matrix, we transform $P^{m,n}$ into a matrix as diagonal as possible
so that the resultant matrix takes the form,
\begin{equation}
\left[
\begin{array}{c|c}
I_r & B\\
\hline
\multicolumn{2}{c}{O}
\end{array}
\right] , 
\label{eq:040216-1909}
\end{equation}
where $I_r$, $O$ and $B$ represents the $r\times r$ identity matrix, an $(N^m-r)\times N^n$ zero matrix and an
$r\times (N^{n}-r)$ matrix, respectively. We can tell that the rank of (\ref{eq:040216-1909}) is $r$. Thus the
procedure for finding 
necessary conditions
for 
the existence of an $M$-dimensional MPSS, 
 is rewritten as follows. 
\begin{enumerate}
 \item Solve equation(\ref{eq:030626-0443}) for $L$ for which one can take $m,n$ such that $L=m+n$ and $N^m > M, N^n > M$.
       
 \item Make the $N^m \times N^n$ matrix $P^{m,n}$ from $\vec{P}_{L}$.

 \item Perform a set of the elementary transformations of $P^{m,n}$ so that the resultant matrix has the form as 
(\ref{eq:040216-1909}).

 \item Find 
necessary conditions
for 
the existence of an $M$-dimensional MPSS, 
from conditions 
that
the rank of the resultant matrix is $M$. 
\end{enumerate}

\subsection{How to construct the matrix-product stationary state}
\label{sec:method-making-MPF}

Suppose that a
necessary condition
obtained by the method in \S\ref{sec:method-how-to-find-the-condition} is
also a sufficient condition for
the existence of an $M$-dimensional MPSS.
Then, how can we find the set of matrices $\{A(0), A(1), \dots , A(N-1), \langle W|, |V\rangle \}$ in
$P^\mathrm{MPF}_L$ (equation(\ref{eq:030828-0005})) from $\vec{P}_{L}$ for a finite number of $L$?  We consider this
problem in \S\ref{sec:try-to-cal-tilders}.  In the subsequent \S\ref{sec:method-how-to-make-Ec-Dc}, we comment on a way
to check whether the obtained matrices can be used for constructing $\vec{P}_{L}$ for any system size $L$ for the case
$N=M$.

\subsubsection{Finding the matrices}
\label{sec:try-to-cal-tilders}

Again we first explain the idea for the case where $N=M=2$.  First thing to do is to solve (\ref{eq:030626-0443}) for
$L=2,3$ and get $\vec{P}_{L=2}$ and $\vec{P}_{L=3}$. If the stationary state $\vec{P}_L$ can be written in the form as
(\ref{eq:030828-0005}), one has  
\begin{align}
\label{eq:031018-0053}
 Z_2 \vec{P}_{L=2} 
 &= 
 \langle W|\binom{E}{D}^{\otimes 2} |V\rangle ,\\
\label{eq:Z_3mpa}
 Z_3 \vec{P}_{L=3} 
 &= 
 \langle W|\binom{E}{D}^{\otimes 3} |V\rangle . 
\end{align}
The matrix form $P^{1,1}$ of (\ref{eq:031018-0053}) is
\begin{equation}
 Z_2 P^{1,1} := \langle W|\binom{E}{D}\left(E~ D\right) |V\rangle .
 \label{eq:031018-0506}
\end{equation}
Let us remember the freedom of the choice of matrices in (\ref{eq:040218-1641}) and (\ref{eq:040218-1642})
by a similarity transformation. If we introduce
\begin{equation}
\begin{split}
 \langle W|S      &=: \langle \widetilde{W}|, \\
 S^{-1} E S       &=: \widetilde{E}, \\
 S^{-1} D S       &=: \widetilde{D}, \\
 S^{-1} |V\rangle &=: |\widetilde{V} \rangle ,
\end{split}
\label{eq:031024-0032}
\end{equation}
with $S$ an invertible 2$\times 2$ matrix, $Z_2 P^{1,1}$ in (\ref{eq:031018-0506}) can also be expressed as
\begin{equation}
 Z_2 P^{1,1} = \langle \widetilde{W}|\binom{\widetilde{E}}{\widetilde{D}}\left(\widetilde{E}~ \widetilde{D}\right) |\widetilde{V}\rangle .
\label{eq:031024-0023}
\end{equation}
Similarly for $L=3$, one has
\begin{equation}
 Z_3
 P^{1,2} = \langle \widetilde{W}|\binom{\widetilde{E}}{\widetilde{D}}\left(\widetilde{E}~ \widetilde{D}\right)^{\otimes 2} |\widetilde{V}\rangle .
\label{eq:Z2_12}
\end{equation}

The key of our methods is to choose $S$ in (\ref{eq:031024-0032})-(\ref{eq:Z2_12}) to be
\begin{equation}
 S :=\left ( E\v~  D\v \right ).
\label{eq:031018-0233}
\end{equation}
In this equation, both $E\v$ and $D\v$ are two-dimensional column vectors, so that this $S$ is a $2 \times 2$ matrix. For
this special choice of $S$, the following holds: 
\begin{equation}
\begin{bmatrix}
 1 & 0\\
 0 & 1
\end{bmatrix} 
=
\left( \widetilde{E}|\widetilde{V}\rangle~ \widetilde{D}|\widetilde{V}\rangle \right) .
\label{eq:031116-1152}
\end{equation}
Using this equation(\ref{eq:031116-1152}) and (\ref{eq:031024-0023}), one has
\begin{equation}
 Z_2 P^{1,1}  = \binom{\langle \widetilde{W}|\widetilde{E}}{\langle \widetilde{W}|\widetilde{D}}.
\label{eq:031207-1740}
\end{equation}
It also follows from (\ref{eq:031024-0032}) that 
\begin{equation}
 \left(E~ D \right)S = S\left(\widetilde{E}~  \widetilde{D} \right).
 \label{eq:031018-0249}
\end{equation}

By virtue of this $S$ one has
\begin{align}
 &           \left(E~   D   \right)\otimes \left(E~   D   \right)        \v \notag\\
=&   \Bigl(E \left(E~   D   \right)\ \   D \left(E~   D   \right) \Bigr)\v \notag\\ 
=&   \Bigl(E \left(E\v~ D\v \right)\ \   D \left(E\v~ D\v \right) \Bigr)   \notag\\ 
=&   \Bigl(E S~~                         D S              \Bigr)         \ (\because \text{equation(\ref{eq:031018-0233})})  \notag\\
=& S \left(  \widetilde{E}~~               \widetilde{D}  \right)\ (\because \text{equation(\ref{eq:031018-0249})}) .
\end{align}
Namely,
\begin{equation}
  \left(E~  D \right)^{\otimes 2}\v = \left(E~  D\right)\v \left(\widetilde{E}~~  \widetilde{D}\right) .
\label{eq:030908-0608}
\end{equation}
From this equation(\ref{eq:030908-0608}), we can obtain
\begin{equation}
  \w \binom{E}{D}^{\otimes k}\left(E~  D \right)^{\otimes (j+1)}\v 
= \w \binom{E}{D}^{\otimes k}\left(E~  D \right)^{\otimes j}\v
 \left(\widetilde{E}~  \widetilde{D}\right) 
\label{eq:030908-0609}
\end{equation}
for $k,j = 1,2,3,\dots$.

An application of (\ref{eq:030908-0609}) for $k=j=1$ leads to  
\begin{equation}
Z_3 P^{1,2} = Z_2 P^{1,1}\left(\widetilde{E}~  \widetilde{D}\right) ,
\label{eq:031018-0308}
\end{equation}
from which one obtains
\begin{align}
 \widetilde{E} &= (Z_2 P^{1,1})^{-1} (Z_3 P^{1,2}[1:2,1:2]),
 \label{eq:031018-1631}\\
 \widetilde{D} &= (Z_2 P^{1,1})^{-1} (Z_3 P^{1,2}[1:2,3:4]) .
 \label{eq:031018-1630}
\end{align}
Here, we introduced the notation $A[b:c\ ,\ d:e]$ for a submatrix of a
matrix $A$ constructed by selecting the row range from the $b$-th
row to the $c$-th row and the column range from the $d$-th column to the $e$-th column.  Now remember that $Z_3 P^{1,2}$
and $Z_2 P^{1,1}$ of these equations can be obtained by solving 
equation
(\ref{eq:030626-0443}) for a small size
$L=2,3$. Therefore we can obtain $\left(\widetilde{E}~ \widetilde{D}\right)$ from the solutions for $L=2,3$.

It should be noted that we need not an explicit expression of $S$ in calculating $\{\widetilde{E},\widetilde{D} \}$
(equations(\ref{eq:031018-1631}) and (\ref{eq:031018-1630})).  $S$ appears only in the equations from which
equations(\ref{eq:031018-1631}) and (\ref{eq:031018-1630}) are derived.

The case $2=N<M$ is treated as follows. Instead of (\ref{eq:031018-0233}), we define the $M \times M$ matrix $S$ as
\begin{equation}
 S := \left(E ~  D \right)^{\otimes \ell} |V\rangle U ,
 \label{eq:031024-0624}
\end{equation}
where $\ell$ is an integer which satisfies $2^\ell\ge M$.  $U$ is a $2^\ell \times M$ matrix by which we choose mutually
independent $M$ column vectors among $2^\ell$ column vectors of a matrix $\left(E ~ D \right)^{\otimes \ell}|V\rangle$
so that there exists $S^{-1}$.  It is noted that when $M=2^\ell$($\ell=2,3,4,\dots$) and $M(=2^\ell)$ column vectors of
a matrix $\left(E ~ D \right)^{\otimes\ell}|V\rangle$ are mutually independent, we can set $U$ as the $M \times M$
identity matrix.

It sometimes really happens that $2^\ell$ column vectors of a matrix $\left(E ~ D \right)^{\otimes \ell}|V\rangle$ are
NOT mutually independent. One of such cases is the case of the 4 dimensional set of matrices for the asymmetric simple
exclusion process(ASEP) (\S\ref{sec:example-ASEP}).  In this case of
$M=2^2$ we can not use
equation(\ref{eq:031024-0624}) with $\ell =2$ and $U=\left(\text{the four dimensional identity matrix}\right)$. This is
because only three in all four column vectors of $\left(E ~ D \right)^{\otimes \ell}|V\rangle$ are mutually independent.
This stems from (\ref{eq:030918-0313}) and the existence of the algebraic relation (\ref{eq:030918-0227}).  For details,
please see the last paragraph of \S\ref{sec:candidates-of-ASEP}.

We should note that equation(\ref{eq:031018-0249}) is also satisfied in this case.  Hence, as a generalisation of
(\ref{eq:030908-0608}) one has
\begin{equation}
\left(E~ D\right)^{\otimes (j+1)} \v U
=
\left(E~ D\right)^{\otimes j}\v U\left(\widetilde{E}~ \widetilde{D}\right) 
\end{equation}
for $j=\ell, \ell+1, \ell+2, \dots $ . Therefore 
\begin{equation}
T\w\binom{E}{D}^{\otimes k}\left(E~ D\right)^{\otimes (j+1)} \v U
=
T\w\binom{E}{D}^{\otimes k}\left(E~ D\right)^{\otimes j}\v U\left(\widetilde{E}~ \widetilde{D}\right) ,
\label{eq:031024-0753}
\end{equation}
that is,
\begin{equation}
T Z_{k+j+1} P^{k,j+1} U = T Z_{k+j} P^{k,j} U\left(\widetilde{E}~ \widetilde{D}\right) 
\label{eq:031024-0810}
\end{equation}
for $j=\ell, \ell+1, \ell+2, \dots $
and $k=1, 2, 3, \dots$
. Here $T$ is a $M \times 2^k$ matrix by which we choose mutually independent $M$ row
vectors among $2^k$ row vectors of a matrix $Z_{k+j} P^{k,j}U$ in the RHS of (\ref{eq:031024-0810}) so that we can solve
about $\widetilde{E}$ and $\widetilde{D}$ in (\ref{eq:031024-0810}) like (\ref{eq:031018-1631}) and
(\ref{eq:031018-1630}). Equation(\ref{eq:031024-0810}) is the generalisation of (\ref{eq:031018-0308}).

Generalisation to the case $2<N\le M$ is straightforward. We define $S$ by
\begin{equation}
 S :=  \Bigl( A(0) ~ A(1) ~ \dots~ A(N-1) \Bigr)^{\otimes \ell}\v U,
\label{eq:040219-2304}
\end{equation}
where $U$ is a $N^\ell \times M$ matrix.
In this case, from (\ref{eq:040218-1641}), we have
\begin{equation}
I_M = \Bigl( \widetilde{A}(0) \ \widetilde{A}(1) \ \dots \widetilde{A}(N-1) \Bigr)^{\otimes \ell}|\widetilde{V}\rangle U ,
\label{eq:040219-2322}
\end{equation}
where $I_M$ is the $M$-dimensional identity matrix. This generalises the equation(\ref{eq:031116-1152}).
If $N=M$ and all column vectors in $\Bigl( A(0) ~ A(1) ~ \dots~ A(N-1) \Bigr)\v $ are mutually independent, 
then we can choose $\ell=1$ and $U=\left(\text{the identity matrix}\right)$ in 
(\ref{eq:040219-2304}) and (\ref{eq:040219-2322}). Namely, they are simplified into the following equations: 
\begin{equation}
S :=  \Bigl( A(0) ~ A(1) ~ \dots~ A(N-1) \Bigr)\v 
\label{eq:031126-0941}
\end{equation}
and
\begin{equation}
I_N = \Bigl( \widetilde{A}(0) \ \widetilde{A}(1) \ \dots \widetilde{A}(N-1) \Bigr)|\widetilde{V}\rangle ,
\label{eq:031126-0939}
\end{equation}
respectively.

\subsubsection{Validity of the obtained matrices for arbitrary size $L$ }
\label{sec:method-how-to-make-Ec-Dc}

Now that we have a set of matrices
$\{\widetilde{A}(0),\widetilde{A}(1),\dots, \widetilde{A}(N-1), |\widetilde{V}\rangle,\langle\widetilde{W}|\}$
by the method described in \S\ref{sec:try-to-cal-tilders}, we would like to know whether the obtained set of matrices is
valid for \textit{any} system size $L$ or not.  It would not be difficult to check this for each small values of $L$
using a computer, but this does not guarantee its validity for general $L$.

In some cases, however, one can check the validity of the obtained set of matrices for any $L$.  In the following
discussions we assume that the following two conditions are satisfied:
\begin{itemize}
 \item $N=M$ and all column vectors in $\Bigl( A(0) ~ A(1) ~ \dots~ A(N-1) \Bigr)\v $ are mutually independent, 
       where we can adopt equation(\ref{eq:031126-0941}) as $S$.

 \item The total hamiltonian $H$ has the form
       \begin{equation}
        H = h^{\mathrm{(L)}} + \sum_{i=1}^{L-1} h_i + h^{\mathrm{(R)}}.
         \label{eq:031210-1852}
       \end{equation}
       Let $I$ denote the 
       $N$-dimensional 
       identity matrix. Then each term in (\ref{eq:031210-1852}) takes the form,
       \begin{equation}
        h_i := I^{\otimes (i-1)} \otimes h_\mathrm{int}\otimes 
        I^{\otimes (L-i-1)}, 
         \label{eq:031210-1853}
       \end{equation}
       \begin{equation}
        h^{\mathrm{(L)}} :=  h^{\mathrm{(\ell)}} \otimes I^{\otimes (L-1)},
         \label{eq:031210-1854}
       \end{equation}
       \begin{equation}
        h^{\mathrm{(R)}} :=   I ^{\otimes (L-1)} \otimes  h^{\mathrm{(r)}} .
         \label{eq:031210-1855}
       \end{equation}
\end{itemize}

Now suppose that we have another set of matrices $\{A_\mathrm{c}(0), A_\mathrm{c}(1),\dots, A_\mathrm{c}(N-1)\}$ which
satisfies the following three equations: 
\begin{equation}
 h_\mathrm{int}\left[
\widetilde{\mathbf{A}}
^{\otimes 2}
 \right]
 = 
\mathbf{A}_\mathrm{c}
\otimes
\widetilde{\mathbf{A}}
-
\widetilde{\mathbf{A}}
\otimes
\mathbf{A}_\mathrm{c},
\label{eq:030829-1146}
\end{equation}
\begin{equation}
 \tildeW 
 \left[h^\mathrm{(\ell)}
\widetilde{\mathbf{A}}
                \right]
 = -\tildeW
\mathbf{A}_\mathrm{c},
\label{eq:030829-1145}
\end{equation}
\begin{equation}
\left[
 h^\mathrm{(r)}
\widetilde{\mathbf{A}}
\right]
\tilde_V
 = 
\mathbf{A}_\mathrm{c}
\tilde_V  ,
\label{eq:030829-1144}
\end{equation}
where $\widetilde{\mathbf{A}}$ and $\mathbf{A}_\mathrm{c}$ are defined as
\begin{equation}
\widetilde{\mathbf{A}} := 
\begin{pmatrix}
\widetilde{A}(0)\\
\widetilde{A}(1)\\
 \vdots\\
\widetilde{A}(N-1) 
\end{pmatrix}   
\quad \text{and} \quad 
\mathbf{A}_\mathrm{c} :=
 \begin{pmatrix}
A_\mathrm{c}(0)\\
A_\mathrm{c}(1)\\
 \vdots\\
A_\mathrm{c}(N-1)
\end{pmatrix} ,
\end{equation}
respectively.
Then we can show that the MPF constructed from the set of matrices
$\{\widetilde{A}(0),\widetilde{A}(1),\dots,\widetilde{A}(N-1), |\widetilde{V}\rangle ,\langle\widetilde{W}|\}$ according
to (\ref{eq:030828-0005}) solves equation(\ref{eq:030626-0443}) for
\textit{any} size $L$. This is due to the so-called cancellation
mechanism\cite{cancel-mech-1,cancel-mech-2,HSP};
all terms in 
$H \vec{P}^\mathrm{MPF}_L = h^{\mathrm{(L)}}\vec{P}^\mathrm{MPF}_L + \sum_{i=1}^{L-1} h_i\vec{P}^\mathrm{MPF}_L + h^{\mathrm{(R)}}\vec{P}^\mathrm{MPF}_L$
are cancelled out by (\ref{eq:030829-1146})-(\ref{eq:030829-1144}).

A good property of our set of matrices
$\{\widetilde{A}(0), \widetilde{A}(1),\dots,\widetilde{A}(N-1), \langle\widetilde{W}|, |\widetilde{V}\rangle \}$ 
enables
us to compute easily a candidate for $\{A_\mathrm{c}(0), A_\mathrm{c}(1),\dots,A_\mathrm{c}(N-1) \}$ in the case
$N=M$. Namely, we can calculate $\{A_\mathrm{c}(0), A_\mathrm{c}(1),\dots,A_\mathrm{c}(N-1)\}$ from the formula 
\begin{equation}
\mathbf{A}_\mathrm{c}
  =
 \begin{pmatrix}
  \widetilde{A}(0)\  {}^t h^\mathrm{(r)} \\
  \widetilde{A}(1)\  {}^t h^\mathrm{(r)} \\
  \vdots\\
  \widetilde{A}(N-1)\  {}^t h^\mathrm{(r)} 
 \end{pmatrix}
  + \Xi
  \left\{
   h_\mathrm{int}
   \left[
\widetilde{\mathbf{A}}
^{\otimes 2}
   \right]
   \right\} 
|\widetilde{V}\rangle ,
\label{eq:030902-2008}
\end{equation}
where we denote a transpose of a matrix $A$, by ${}^t A$ and $\Xi\{\cdot\}$ is the operator which transforms 
an $N^2$-dimensional vector ${}^t \left( B_1, B_2,\dots,B_{N^2} \right)=:\mathbf{B}[N^2]$
into 
\begin{equation*}
 \begin{bmatrix}
      B_1 &     B_2 & \dots  & B_N    \\
  B_{N+1} & B_{N+2} & \dots  & B_{2N}  \\
  B_{2N+1}& \vdots  & \vdots & \vdots \\
 \vdots   & \vdots  & \vdots & \vdots \\
  \dots   & \dots   & \dots  & B_{N^2}
 \end{bmatrix}.
\end{equation*}
This operator satisfies
$
\Xi\left\{
\mathbf{B}[N^2]+\mathbf{C}[N^2]
\right\}=
\Xi\left\{
\mathbf{B}[N^2]
\right\}+
\Xi\left\{
\mathbf{C}[N^2]
\right\} 
$
and
$
 \Xi\left\{
\mathbf{B}[N]
            \otimes
\mathbf{C}[N]
           \right\}
           =
\mathbf{B}[N]
           \otimes
{}^t \mathbf{C}[N].
$
It is noted that the argument of the LHS of the latter is a vector with $N^2$ components.

Derivation of (\ref{eq:030902-2008}) is as follows.  By virtue of (\ref{eq:031126-0939}) (the good property of our set
of matrices), we can derive from  (\ref{eq:030829-1144})
\begin{equation}
 {}^t h^\mathrm{(r)}
 = 
 {}^t \mathbf{A}_\mathrm{c}
|\widetilde{V}\rangle. 
\label{eq:030902-1959}
\end{equation}
We perform $\Xi\{\text{ equation(\ref{eq:030829-1146})}\} |\widetilde{V}\rangle$ to get
\begin{equation}
 \Xi
  \left\{
   h_\mathrm{int}
   \left[
\widetilde{\mathbf{A}}
    \otimes
\widetilde{\mathbf{A}}
  \right]
 \right\}
|\widetilde{V}\rangle
= 
\left[
\mathbf{A}_\mathrm{c}
 \otimes
 {}^t \widetilde{\mathbf{A}}
\right]
|\widetilde{V}\rangle
-
\left[
\widetilde{\mathbf{A}}
 \otimes
 {}^t \mathbf{A}_\mathrm{c}
\right]
|\widetilde{V}\rangle .
\label{eq:031116-1334}
\end{equation}
Using (\ref{eq:031126-0939}), (\ref{eq:030902-1959}) and a relation 
$\left(A\otimes B\right) |\widetilde{V}\rangle = A\otimes \left(B |\widetilde{V}\rangle \right)$, 
the RHS of (\ref{eq:031116-1334}) is 
\begin{equation}
\mathbf{A}_\mathrm{c}
 \otimes
\left[
 {}^t \widetilde{\mathbf{A}}
|\widetilde{V}\rangle
\right]
 - 
\widetilde{\mathbf{A}}
\otimes
\left[
 {}^t \mathbf{A}_\mathrm{c}
|\widetilde{V}\rangle
\right] 
=
\mathbf{A}_\mathrm{c}
 \otimes
I_N
-
\widetilde{\mathbf{A}}
 \otimes
 {}^t h^\mathrm{(r)} .
\end{equation}
This results in the formula (\ref{eq:030902-2008}).

Once a candidate for the set of matrices 
$\{ \widetilde{\mathbf{A}} , \mathbf{A}_\mathrm{c} , |\widetilde{V}\rangle , \langle\widetilde{W}| \}$
is obtained from (\ref{eq:030902-2008}), we must check whether the candidate satisfies
equations(\ref{eq:030829-1146})-(\ref{eq:030829-1144}). If these equations hold,
it means that the obtained set of matrices is valid for any $L$.

So far we have not succeeded in extending the formula (\ref{eq:030902-2008}) to the case $N\ne M$.  This is an open question.

\section{Examples}
\label{sec:ex}

In this section, the application of our methods in \S\ref{sec:method} is illustrated by performing explicit calculations for three models:
\begin{enumerate}

 \item the asymmetric exclusion process(ASEP)
 \cite{review-of-matrix-product-solution,EsslerRittenberg,%
MallickSandow1997,Sasamoto-Polynomial-approach-1,Sasamoto-Polynomial-approach-2,B-Polynomial-approach,SS,USW}
 (\S\ref{sec:example-ASEP}), 

 \item the model studied in \cite{J2}(\S\ref{sec:J-model-031127-0852}),

 \item  a hybrid of the model(1) and the model(2)(\S\ref{seq:example-hybrid}). 
\end{enumerate}
For the first two models, it is already known that there exist 
MPSSs for special values of the model parameters. We will see that the
conditions and the matrices can be reproduced from our methods.
For the third model, we apply the same methods and find a MPSS, 
which was unknown.


\subsection{The asymmetric simple exclusion process (ASEP)}
\label{sec:example-ASEP}

In this subsection, we treat the asymmetric simple exclusion 
process(ASEP). The model is defined on the one-dimensional lattice
whose size is $L$.  Each site can take two states: a site is either empty or occupied by a particle. The time of the
model is
continuous one
(i.e. the random sequential update).
A particle in the bulk hops to the 
left (resp. right) neighbour site with a rate
$p_\mathrm{L}$ (resp. $p_\mathrm{R}$) if the left (resp. right) 
neighbour site is empty. A particle at the rightmost site is removed with
a rate $\beta$ if the site is occupied by a particle.  At the leftmost site, a particle is injected with a rate $\alpha$
if the site is empty.

The total hamiltonian $H$ of this model has the form in (\ref{eq:031210-1852})-(\ref{eq:031210-1855}).  For the case of
the ASEP, $h^{\mathrm{(\ell)}}$ of (\ref{eq:031210-1854}) and $h^{\mathrm{(r)}}$ of (\ref{eq:031210-1855}), which
express the left and right boundary condition respectively, are
\begin{equation}
 h^{\mathrm{(\ell)}} := 
 \left[ \begin {array}{cc} \alpha&0\\\noalign{\medskip}-\alpha&0\end {array} \right] 
\quad\text{and}\quad
  h^{\mathrm{(r)}} := 
 \left[ \begin {array}{cc} 0&-\beta\\\noalign{\medskip}0&\beta\end {array} \right] ,
\label{eq:030918-0315}
\end{equation}
in a basis of states whose order is $\left(|\emptyset\rangle,|\mathrm{A}\rangle\right)$.  The interaction hamiltonian
$h_\mathrm{int}$ of (\ref{eq:031210-1853}) is defined as 
\begin{equation}
 h_\mathrm{int} :=
 \left[ \begin {array}{cccc} 0&0&0&0\\\noalign{\medskip}0&q&-1&0\\\noalign{\medskip}0&-q&1&0\\\noalign{\medskip}0&0&0&0\end {array} \right] ,
\label{eq:031212-0936}
\end{equation}
in a basis of states whose order is
$\left(|\emptyset\emptyset\rangle,|\emptyset\mathrm{A}\rangle,|\mathrm{A}\emptyset\rangle,|\mathrm{A}\mathrm{A}\rangle\right)$. 
In (\ref{eq:031212-0936}), we have introduced the parameter $q:=p_\mathrm{L}/p_\mathrm{R}$ and set $p_\mathrm{R}=1$.
In the following, we assume that $\alpha > 0, \beta > 0, q>0$.

\subsubsection{Determination of the 
necessary condition for
the existence of an $M$-dimensional MPSS
 ---a case of the ASEP---}
\label{sec:candidates-of-ASEP}

Let us find
a necessary condition for the existence of the $M(=1,2)$-dimensional set of matrices $\{E,D,\v,\w\}$.
We perform the following calculations 
according to the procedure in
\S\ref{sec:method-how-to-find-the-condition}.  First, we obtain $P^{2,2}$(for its definition, see
equation(\ref{eq:031212-0957})) by solving equation(\ref{eq:030626-0443}) for the $(L=)4$-site system.  Next we perform
a set of elementary transformations so that the resultant matrix has 
the form in (\ref{eq:040216-1909}).  For the
present case, the resultant matrix is
\begin{align}
 \begin{cases}
  \left[
    \begin{array}{cccc}
     1&0&0&0\\
     \noalign{\medskip}0&1&0&-{\frac {q+\beta}{\beta}}\\
     \noalign{\medskip}0&0&1&{\beta}^{-1}\\
     \noalign{\medskip}0&0&0&0
    \end{array}
  \right] 
  & ( f_\mathrm{ASEP1} \ne 0 \text{\ and\ }  f_\mathrm{ASEP2} \ne 0)\cr
  \left[
    \begin{array}{cccc}
     1&-{\frac {-1+\beta+q}{\beta}}&-{\frac {-1+\beta+q}{\beta}}&{\frac { \left( -1+\beta+q \right) ^{2}}{{\beta}^{2}}}\\
     \noalign{\medskip}0&0&0&0\\
     \noalign{\medskip}0&0&0&0\\
     \noalign{\medskip}0&0&0&0
    \end{array}
  \right] 
  & ( f_\mathrm{ASEP1} = 0 )\cr
  \left[
    \begin{array}{cccc}
     1&
     0&
     -{\frac { \left( -1+\beta+q \right) ^{2}q}{\beta\, \left( q+\beta \right) }}&
     -{\frac { \left( -1+\beta+q \right) ^{2}q}{ \left(q+\beta \right) {\beta}^{2}}}\\
     \noalign{\medskip}0&
     1&
     -{\frac {-2\,q-\beta+{q}^{2}+\beta\,q}{q+\beta}}&
     -{\frac { \left( -1+\beta+q \right)  \left( 2\,q+\beta \right) }{\beta\, \left( q+\beta \right) }}\\
     \noalign{\medskip}0&
     0&
     0&
     0\\
     \noalign{\medskip}0&0&0&0
    \end{array}
 \right] 
  & ( f_\mathrm{ASEP1} \ne 0 \text{\ and\ }  f_\mathrm{ASEP2} = 0)\cr
 \end{cases}
\label{eq:040217-1421}
\end{align}
where
$
 f_\mathrm{ASEP1} := q +\alpha+\beta -1
\quad\text{and}\quad
 f_\mathrm{ASEP2} := 
 {q}^{2}+q\left( \alpha +\beta-1\right) +\alpha \beta .
$
From (\ref{eq:040217-1421}), we can see that the rank of $P^{2,2}$ is 1 when $ f_\mathrm{ASEP1}=0 $.  So
$f_\mathrm{ASEP1} = 0$ is the 
necessary condition
for 
the existence of an $M(=1)$-dimensional MPSS.
According to
\cite{EsslerRittenberg,MallickSandow1997,Sasamoto-Polynomial-approach-1,Sasamoto-Polynomial-approach-2},
this condition is the true one.  We can also
see that the rank of $P^{2,2}$ is 2 when $ f_\mathrm{ASEP2}=0$ and $f_\mathrm{ASEP1}\ne 0$ from (\ref{eq:040217-1421}).
So this condition is the 
necessary condition
for 
the existence of an $M(=2)$-dimensional MPSS.
This is also true according to the
reference\cite{EsslerRittenberg,MallickSandow1997,Sasamoto-Polynomial-approach-1,Sasamoto-Polynomial-approach-2}.  In the next
\S\ref{sec:find-MPF-of-ASEP}, we try to find a two-dimensional set of matrices by using $f_\mathrm{ASEP2} = 0$.  That
is, we use
\begin{equation}
 \alpha = g_\mathrm{A}(q,\beta) , 
\quad\text{where}\quad
g_\mathrm{A}(q,\beta) :=
 \frac{
 -q(q-1+\beta)
}{
q+\beta
}.
\label{eq:031212-1214}
\end{equation}

Here we describe the rank deficiency which happens for the ASEP
due to the existence of the algebraic relations, such as
\begin{equation}
 DE - q ED \propto (E+D)
\label{eq:030918-0227}
\end{equation}
and 
\begin{equation}
 D|V\rangle \propto |V\rangle.
\label{eq:030918-0313}
\end{equation}
These equations can be shown to be satisfied using
(\ref{eq:031212-0936}) , (\ref{eq:030829-1146}) , 
 (\ref{eq:030829-1144}),
(\ref{eq:030918-0315}) and (\ref{eq:031212-1735}).
Because of (\ref{eq:030918-0227})
and (\ref{eq:030918-0313}), the rank of $P^{2,2}$ is at most 3
(cf. (\ref{eq:040217-1421})). More generally, the rank of  
$
 \langle W|\binom{E}{D}^{\otimes k}(E~  D)^{\otimes 2}|V\rangle
 \quad (k=2,3,4,\dots)
$
is also at most 3. 
It should be noted that, generally, not only the existence of an MPSS but also an algebraic relation between matrices in
the MPSS could cause a rank deficiency. 

\subsubsection{Construction of the $(M=)2$-dimensional set of matrices---a case of the ASEP---}
\label{sec:find-MPF-of-ASEP}

Let us find the $(M=)2$-dimensional representation of the MPSS of the ASEP. In the following, we use expressions for
which we have eliminated $\alpha$ by using (\ref{eq:031212-1214})
.

Firstly, we calculate $\{\widetilde{E}, \widetilde{D}\}$. We solve equation(\ref{eq:030626-0443}) for $L=2$ and
transform the solution into the matrix form $P^{1,1}$
:
\begin{equation*}
 Z_2 P^{1,1}=
\begin{bmatrix}
 1 &
{\frac {{g_\mathrm{A}(q,\beta)}}{\beta}}\\
{\frac { \left( q+q\beta+{\beta}^{2} \right) {g_\mathrm{A}(q,\beta)}}{\beta\, \left(
 q+\beta \right) }} &
{\frac {{{g_\mathrm{A}(q,\beta)}}^{2}}{{\beta}^{2}}}
\end{bmatrix}.
\end{equation*}
From this equation, we define $\langle W_1|$ and $\langle W_2|$ by
\begin{equation}
  Z_2 P^{1,1}
=:
\binom{\langle {W}_1|}{\langle {W}_2|}.
\label{eq:031207-1754}
\end{equation}
We solve equation(\ref{eq:030626-0443}) also for $L=3$ and transform the solution into the matrix form $P^{1,2}$:
\begin{equation*}
Z_3 P^{1,2} =
\begin{bmatrix}
1 &
 {\frac {{g_\mathrm{A}(q,\beta)}}{\beta}}
&
 {\frac { \left( q+q\beta+{\beta}^{2} \right) {g_\mathrm{A}(q,\beta)}}{\beta\, \left( q+\beta \right) }}
&
{\frac {{{g_\mathrm{A}(q,\beta)}}^{2}}{{\beta}^{2}}}
\\
 {\frac { \left( {q}^{2}+2\,{q}^{2}\beta+3\,q{\beta}^{2}+{\beta}^{3} \right) {g_\mathrm{A}(q,\beta)}}{ \left( q+\beta \right) ^{2}\beta}}
&
{\frac { \left( q+q\beta+{\beta}^{2} \right) {{g_\mathrm{A}(q,\beta)}}^{2}}{ \left( q+\beta \right) {\beta}^{2}}}
&
 {\frac { \left( {q}^{2}\beta+2\,q{\beta}^{2}+q+{\beta}^{3} \right) {{g_\mathrm{A}(q,\beta)}}^{2}}{ \left( q+\beta \right) {\beta}^{2}}}
&
 {\frac {{{g_\mathrm{A}(q,\beta)}}^{3}}{{\beta}^{3}}}
\end{bmatrix}.
\end{equation*}
From (\ref{eq:031018-1631}) and (\ref{eq:031018-1630}), we can obtain $\widetilde{E}$ and $\widetilde{D}$:
\begin{equation*}
 \widetilde{E} =
\begin{bmatrix}
{\frac {2\,q+\beta}{q+\beta}}&
\frac{g_\mathrm{A}(q,\beta)}{\beta}\\
{\frac {\beta }{q-1+\beta}}&
0
\end{bmatrix}
\quad\text{and}\quad
 \widetilde{D} =
\frac{g_\mathrm{A}(q,\beta)}{\beta}
\begin{bmatrix}
q+\beta &
0\\
\beta &
1
\end{bmatrix}.
\end{equation*}

Secondly, we calculate $\{\langle\widetilde{W}|, |\widetilde{V}\rangle\}$. For $|\widetilde{V}\rangle $, 
we can use 
\begin{equation}
\binom{1}{0}  = \widetilde{E}|\widetilde{V}\rangle ,
\label{eq:031207-2045}
\end{equation}
or 
\begin{equation}
\binom{0}{1}  = \widetilde{D}|\widetilde{V}\rangle
\label{eq:031207-2044}
\end{equation}
which are the left and right half of (\ref{eq:031116-1152}), respectively.
If $\det(\widetilde{E})=q/(q+\beta)\ne 0$, we can obtain $|\widetilde{V}\rangle$ by 
$(\widetilde{E})^{-1}\binom{1}{0}$. Or, if $\det(\widetilde{D})=q^2(q-1+\beta)^2/(\beta^2(q+\beta))\ne 0 $, we can obtain
$|\widetilde{V}\rangle$ by $(\widetilde{D})^{-1}\binom{0}{1}$: 
$
|\widetilde{V}\rangle =  \left(\widetilde{E}\right)^{-1}\binom{1}{0}
=  \left(\widetilde{D}\right)^{-1}\binom{0}{1}
= \binom{0}{\frac{\beta}{g_\mathrm{A}(q,\beta)}}.
$
Under the same condition, we can get $\langle\widetilde{W}|$ by
\begin{equation}
 \langle W_1| = \langle\widetilde{W}|\widetilde{E},
\label{eq:031207-2342}
\end{equation}
or
\begin{equation}
 \langle W_2| = \langle\widetilde{W}|\widetilde{D},
\label{eq:031207-2341}
\end{equation}
which can be shown from (\ref{eq:031207-1754}) and (\ref{eq:031207-1740}).
Thus
$
\langle\widetilde{W}| = \langle W_1| (\widetilde{E})^{-1}
 = \langle W_2| (\widetilde{D})^{-1}
 = \left(1 , \frac{g_\mathrm{A}(q,\beta)}{\beta}  \right).
$

At this point, we should check that $\vec{P}^\mathrm{MPF}$
(see (\ref{eq:030828-0005})) calculated by our
\\
$\{\widetilde{E},\widetilde{D},\tildeW,\tilde_V\}$ is also the solution of (\ref{eq:030626-0443}) for any size
$L$. For this check, it is adequate that there exist $E_\mathrm{c}$ and $D_\mathrm{c}$ which satisfy
equations(\ref{eq:030829-1146})-(\ref{eq:030829-1144}). 
Using the above-mentioned result, we obtain 
a candidate
for
$\{E_\mathrm{c}, D_\mathrm{c} \}$ by (\ref{eq:030902-2008})
\begin{equation}
 -E_\mathrm{c} = D_\mathrm{c} = g_\mathrm{A}(q,\beta)
  \begin{bmatrix}
   1 & 0\\
   0 & 1
  \end{bmatrix}.
\label{eq:031212-1735}
\end{equation}
It is easy to show this $\{E_\mathrm{c}, D_\mathrm{c} \}$ satisfies 
equations(\ref{eq:030829-1146})-(\ref{eq:030829-1144}).
Therefore, it is not only a candidate but also the true $\{E_\mathrm{c} , D_\mathrm{c} \}$.

We comment on the relation between the known set of matrices and ours.
The known two-dimensional matrices, 
$\{E,D\}$\cite{EsslerRittenberg,MallickSandow1997,Sasamoto-Polynomial-approach-1,Sasamoto-Polynomial-approach-2}, which we denote
$\{E_\mathrm{A} , D_\mathrm{A} \}$, are given by
\begin{equation*}
 E_\mathrm{A} := 
\begin{bmatrix}
 {\frac {1-q}{{g_\mathrm{A}(q,\beta )}}}
&
0
\\
1
&
(1-q)\left(1+\frac{q}{g_\mathrm{A}(q,\beta)}  \right)
\end{bmatrix}
\quad\text{and}\quad
   D_\mathrm{A} := 
 \begin{bmatrix}
 {\frac {1-q}{\beta}}
&
(1-q)\left(1-\frac{1}{q}\right)
\\
0
&
(1-q)\left( 1+\frac{q}{\beta } \right)
 \end{bmatrix} .
\end{equation*}
The relation between $\{E_\mathrm{A} , D_\mathrm{A} \}$ and $\{\widetilde{E}, \widetilde{D}\}$ is 
\begin{equation}
 S_\mathrm{A}
 \binom{\widetilde{E}}{\widetilde{D}}
\left(S_\mathrm{A}\right)^{-1}
\frac{(1-q)}{g_\mathrm{A}(q,\beta ) }
=
 \binom{E_\mathrm{A}}{D_\mathrm{A}} , 
\quad\text{where }\quad
 S_\mathrm{A} := 
  c
  \left[
   \begin {array}{cc}
   {\frac { 1-q  }{{g_\mathrm{A}(q,\beta )}}}&
   {\frac { 1-q  }{\beta}}\\
 \noalign{\medskip}1&
  0
  \end {array}
 \right] 
\label{eq:031220-1428}
\end{equation}
and $c$ is a free parameter. This $S_\mathrm{A} $ exists when
$
 \det{S_\mathrm{A}} = -\frac{c^2 (1-q)}{\beta} \ne 0
$
is satisfied. Moreover, our result (\ref{eq:031212-1735}) corresponds to
the result in the
reference\cite{EsslerRittenberg,MallickSandow1997,Sasamoto-Polynomial-approach-1,Sasamoto-Polynomial-approach-2}.  
This is because $\{E_\mathrm{c}^\mathrm{A}, D_\mathrm{c}^\mathrm{A}\}$ defined as
$
\binom{E_\mathrm{c}^\mathrm{A}}{D_\mathrm{c}^\mathrm{A}}
:=
 S_\mathrm{A}
 \binom{E_\mathrm{c}}{D_\mathrm{c}}
\left(S_\mathrm{A}\right)^{-1}
\frac{(1-q)}{g_\mathrm{A}(q,\beta ) }
$
are of type of $\text{(a scalar)}\times\text{(the two-dimensional identity matrix)}$.

Furthermore, we have checked that there exists relations like (\ref{eq:031220-1428}) between our $(M=)3,4$-dimensional
sets of matrices which are obtained by the method in \S\ref{sec:try-to-cal-tilders} and the sets of matrices in the
reference\cite{MallickSandow1997,Sasamoto-Polynomial-approach-1,Sasamoto-Polynomial-approach-2}.

\subsection{The model in \cite{J2}}
\label{sec:J-model-031127-0852}

The model treated in this subsection is the one studied in \cite{J2}.
It is defined on the one-dimensional lattice which has $L$ sites.  
Each site can be either
empty or occupied by a particle. The time evolution of the model is continuous(i.e. the random sequential update).
Between nearest neighbour sites of the chain, there stochastically 
occur three types of processes: diffusion, coagulation
and decoagulation.  The rate of each process is
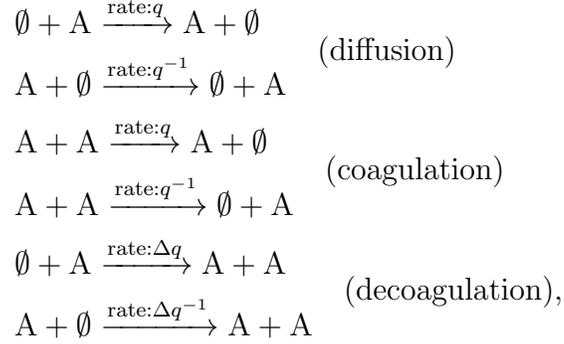
\begin{align}
&
 \begin{aligned}
   \emptyset + \mathrm{A} &\xrightarrow{\text{rate:} q     }  \mathrm{A} + \emptyset \\
   \mathrm{A} + \emptyset &\xrightarrow{\text{rate:} q^{-1}}  \emptyset  + \mathrm{A}
  \end{aligned}
  \text{\quad (diffusion)}
\notag
\\
&
 \begin{aligned}
   \mathrm{A} + \mathrm{A} &\xrightarrow{\text{rate:} q     }  \mathrm{A} + \emptyset \\
   \mathrm{A} + \mathrm{A} &\xrightarrow{\text{rate:} q^{-1}}  \emptyset  + \mathrm{A}
  \end{aligned}
  \text{\quad (coagulation)}
\notag
\\
&
 \begin{aligned}
   \emptyset + \mathrm{A} &\xrightarrow{\text{rate:} \Delta q     }  \mathrm{A} + \mathrm{A} \\
   \mathrm{A} + \emptyset &\xrightarrow{\text{rate:} \Delta q^{-1}}  \mathrm{A} + \mathrm{A}
  \end{aligned}
  \text{\quad (decoagulation)},
\notag
\end{align}
where $\mathrm{A}$ and $\emptyset$ represent a particle and an empty site, respectively. At the leftmost site, a
particle is injected with a rate $\alpha$ if the site is empty and 
removed with a rate $\beta$ if the site is occupied. This model has a
reflective boundary condition at the rightmost site.  In the following, we assume 
$
q > 0,\  \alpha > 0,\  \beta > 0,\  \Delta > 0,
$
although there exists a (four-dimensional) MPSS, which we do not treat
in this paper, also in a case with $\alpha = \beta = 0$\cite{HSP}.

The total hamiltonian $H$ of this model takes the form in (\ref{eq:031210-1852})-(\ref{eq:031210-1855}), where
$h^{\mathrm{(\ell)}}$ and $ h^{\mathrm{(r)}}$ are given by
\begin{equation}
 h^{\mathrm{(\ell)}} 
 :=
 \left[ \begin{array}{cc} \alpha&-\beta \\\noalign
 {\medskip}-\alpha&\beta\end{array} \right]
\quad\text{and}
 \quad
 h^{\mathrm{(r)}} 
 :=
 \left[ \begin{array}{cc} 0 & 0 \\\noalign
 {\medskip} 0 & 0 \end{array} \right] 
\label{eq:040220-0935}
\end{equation}
in a basis of states whose order is
$\left(|\emptyset\rangle,|\mathrm{A}\rangle\right)$,
and the interaction term $h_\mathrm{int}$ is defined as
\begin{equation}
 h_\mathrm{int} :=
 \left[ 
{\begin{array}{cccc}
 0 & 0 &  0  & 0 \\
 [2ex]
0 & (1 + \Delta) \,q &  - {\displaystyle \frac {1}{q}}  &  - {\displaystyle \frac {1}{q}}  \\
 [2ex]
 0  &  - q  &  {\displaystyle \frac {1+\Delta}{q}} &  - q \\
 [2ex]
0 &  - \Delta \,q &  - {\displaystyle \frac {\Delta }{q}}  & q + {\displaystyle \frac {1}{q}} 
\end{array}}
 \right] 
\label{eq:031127-0900}
\end{equation}
in a basis of states whose order is
$\left(|\emptyset\emptyset\rangle,|\emptyset\mathrm{A}\rangle,|\mathrm{A}\emptyset\rangle,|\mathrm{A}\mathrm{A}\rangle\right)$. 
It is noted that the overall signs of (\ref{eq:040220-0935}) and
(\ref{eq:031127-0900}) are opposite to the ones in \cite{J2}.

\subsubsection{Determination of the 
necessary condition for
the existence of an $M$-dimensional MPSS.
   ---a case of the model in \cite{J2}---}
\label{sec:candidates-of-J}

Let us find a 
necessary condition
for 
the existence of the $(M=)2$-dimensional set of matrices $\{E,D,\v,\w\}$.
According to the procedure in
\S\ref{sec:method-how-to-find-the-condition}, we solve
equation(\ref{eq:030626-0443}) for $L=4$ and obtain
the four dimensional square matrix 
$P^{2,2}$(for its definition, see equation(\ref{eq:031212-0957})).  
We perform a set of elementary transformations so that the 
resultant matrix has the form in (\ref{eq:040216-1909}).  
The resultant matrix is
\begin{align}
 \begin{cases}
  I_4 
  & (f_{\mathrm{J}} \ne 0)\cr
  \left[
      \begin {array}{cccc}
      1&0&0&0\\\noalign{\medskip}0&1&{q}^{2}&\Delta\\\noalign{\medskip}0&0&0&0\\\noalign{\medskip}0&0&0&0
      \end {array}
  \right] 
  & (f_{\mathrm{J}}=0) \cr
\end{cases}
\label{eq:040217-1131}
\end{align}
where $I_4$ is the four dimensional identity matrix and 
$
 f_{\mathrm{J}} :=  \Delta\beta q + \Delta - q\alpha - q^2\Delta .
$
In the process of transformation to obtain (\ref{eq:040217-1131}), 
two ``big'' polynomials, each of which has more than 493 terms, 
appear chiefly in factors of numerators of the diagonal
elements. It seems very unlikely that these
polynomials contain important information about the existence of a MPSS.
Hence we assume that these polynomials can not vanish, 
so that we can divide by them. From
(\ref{eq:040217-1131}), we can tell that the rank of $P^{2,2}$ changes
from 4 to 2 when $f_{\mathrm{J}}=0$.  
This condition, $f_{\mathrm{J}}=0$, can be rewritten as
\begin{equation}
 \alpha = g_\mathrm{J}(q,\beta,\Delta),
\quad\text{where}\quad
g_\mathrm{J}(q,\beta,\Delta) :=  (1/q - q + \beta ) \Delta  .
\label{eq:030829-0446}
\end{equation}
In the next \S\ref{sec:find-MPF-of-J}, we assume this condition
(\ref{eq:030829-0446})
holds.

\subsubsection{Construction of the $(M=)2$-dimensional set of matrices
   ---a case of the model in \cite{J2}---}
\label{sec:find-MPF-of-J}

Let us find the $(M=)2$-dimensional representation of the MPSS of the model. 
Because calculations we have to perform here are quite similar to \S\ref{sec:find-MPF-of-ASEP}, we mainly focus on
summarising the results, for which we have eliminated $\alpha$ using (\ref{eq:030829-0446}).
The set of matrices of this model is 
\begin{equation*}
 \widetilde{E}
=
 \left[ 
{\begin{array}{rc}
1 & 0 \\
0 & {\displaystyle \frac {1}{q^{2}}} 
\end{array}}
 \right]
\quad\text{and}\quad
 \widetilde{D}
=
 \left[ 
{\begin{array}{rc}
0 & 0 \\
1 & {\displaystyle \frac {\Delta }{q^{2}}} 
\end{array}}
 \right]
\qquad\left(\text{
if
}\quad
 \Delta \ne \frac{g_\mathrm{J}(q,\beta,\Delta)}{\beta}
\right)
\end{equation*}
\begin{equation*}
\tilde_V 
 = 
\begin{bmatrix}
1\\
0
\end{bmatrix}
\quad\text{and}\quad
\langle \widetilde{W}|=
 \left[ 
1\quad \frac {g_\mathrm{J}(q,\beta,\Delta)}{\beta}
 \right] 
.
\end{equation*}
The candidates for $E_\mathrm{c}$ and $D_\mathrm{c}$ which can be obtained from (\ref{eq:030902-2008}) are  
\begin{equation*}
 E_\mathrm{c} = 
\begin{bmatrix}
 0 & 0 \\
 0 & \frac{\Delta (q^2-1)}{q^3}
\end{bmatrix}
\quad\text{and}\quad
 D_\mathrm{c} = 
-E_\mathrm{c}  .
\end{equation*}
It is easy to check that these also satisfy 
equations(\ref{eq:030829-1146})-(\ref{eq:030829-1144}).
Therefore, they are 
the true $E_\mathrm{c}$ and $D_\mathrm{c}$.

\subsection{A hybrid model}
\label{seq:example-hybrid}

In this subsection, we treat a model similar to the one 
in \S\ref{sec:J-model-031127-0852}. The only differences are
boundary conditions.  The boundary conditions of the model treated here are the same as those of the model in
\S\ref{sec:example-ASEP}. Therefore the total hamiltonian $H$ of this model is defined by
(\ref{eq:031210-1852})-(\ref{eq:031210-1855}) where $h^{\mathrm{(\ell)}} $ and $h^{\mathrm{(r)}}$ are the same
hamiltonians as those defined in (\ref{eq:030918-0315}), and the interaction hamiltonian $h_\mathrm{int}$ is the same as
that of the model in \cite{J2}(except an overall sign) and is given by (\ref{eq:031127-0900}). 
In the following calculations, we assume
$
q > 0,\  \alpha > 0,\  \beta > 0,\  \Delta > 0.
$

\subsubsection{Determination of the 
necessary condition for
the existence of an $M$-dimensional MPSS
 ---a case of a hybrid model---}
\label{sec:candidates-of-hybrid}

Let us find
a
necessary condition for the existence of the $(M=)2$-dimensional set of matrices $\{E,D,\v,\w\}$.
We perform the following calculations as in
\S\ref{sec:candidates-of-ASEP}
according to the procedure in
\S\ref{sec:method-how-to-find-the-condition}.  We solve
equation(\ref{eq:030626-0443}) for $L=4$ and obtain
the four dimensional square matrix $P^{2,2}$. The resultant matrix after a set of
elementary transformations is
\begin{align}
 \begin{cases}
  I_4 
  & (f_\mathrm{hy} \ne 0 )\cr
  \left[ 
    \begin{array}{cccc}
     1&0&0&0\\
     \noalign{\medskip}0&1&q \left( q+\beta \right) &\Delta\\
     \noalign{\medskip}0&0&0&0\\
     \noalign{\medskip}0&0&0&0
    \end{array}
  \right] 
  & (f_\mathrm{hy} = 0)\cr
 \end{cases}
\label{eq:040217-1950}
\end{align}
where $I_4$ is the four dimensional identity matrix and 
$
 f_\mathrm{hy} :=  -\Delta + q^2 \Delta + q \alpha .
$
As in \S\ref{sec:candidates-of-J}, in the process of transforming to (\ref{eq:040217-1950}), two
``big'' polynomials, each of which has more than 501 terms, 
appear chiefly in factors of numerators of the
diagonal elements.  We again assume that these polynomials can not be zero, so
that we can divide by them.  
From (\ref{eq:040217-1950}), we can tell that the rank deficiency
occurs(the change of the rank is 4 to 2) when $f_\mathrm{hy} = 0$.  
We regard the condition, $f_{\mathrm{hy}}=0$, as 
a 
necessary condition
for 
the existence of an $M(=2)$-dimensional MPSS.
  For use in the next \S\ref{sec:find-MPF-of-hybrid}, we transform
$f_{\mathrm{hy}}=0$ into the following form
\begin{equation}
 \alpha = g_\mathrm{hy}(q, \Delta) 
\quad\text{where}\quad
 g_\mathrm{hy}(q, \Delta) := {\frac {\Delta\, \left( 1-{q}^{2} \right) }{q}} .
\label{eq:031213-1554}
\end{equation}

\subsubsection{Construction of the $(M=)2$-dimensional set of matrices---a case of a hybrid model ---}
\label{sec:find-MPF-of-hybrid}

Let us find the $(M=)2$-dimensional representation of the MPSS of the hybrid model. 
Because calculations we have to perform here are quite similar to \S\ref{sec:find-MPF-of-ASEP}, we mainly focus on
summarising the results, for which we have eliminated $\alpha$ by using (\ref{eq:031213-1554}).
The set of matrices of this model is
\begin{equation*}
 \widetilde{E} =
 \left[ \begin {array}{cc} {q}^{2}&0\\\noalign{\medskip}{\frac {\beta\,q}{\Delta}}&1\end {array} \right] 
\quad
 \widetilde{D} =
 \left[ \begin {array}{cc} 0&0\\\noalign{\medskip}q \left( q+\beta \right) &\Delta\end {array} \right]
\quad
\tilde_V =
 \left[ \begin {array}{c} 1/{q}^{2}\\\noalign{\medskip}-{\frac {\beta}{\Delta\,q}}\end {array} \right] 
\quad
\langle \widetilde{W}|=
 \left[ \begin {array}{cc} 1&{\frac {{g_\mathrm{hy}(q, \Delta)}}{\beta}}\end {array} \right]. 
\end{equation*}
And 
candidates
for $E_\mathrm{c}$ and $D_\mathrm{c}$ are
\begin{equation}
 -E_\mathrm{c} = D_\mathrm{c} = 
  \begin{bmatrix}
   0     & 0 \\
   \beta & {g_\mathrm{hy}(q, \Delta)}
  \end{bmatrix} .
\label{eq:031214-0729}
\end{equation}
We have checked that these candidates for 
$E_\mathrm{c}$ and $D_\mathrm{c}$ satisfy
equations(\ref{eq:030829-1146})-(\ref{eq:030829-1144}),
so that 
(\ref{eq:031214-0729}) represents the true 
$E_\mathrm{c}$ and $D_\mathrm{c}$.

\section{Summary}
\label{sec:summary}

In this paper, we have given a systematic way to find and construct exact finite dimensional matrix product stationary
states for one-dimensional 
stochastic models which have $N$ states per site.
More precisely, we have explained (\S\ref{sec:method}): 
\begin{enumerate}
 \item a systematic way to search 
necessary conditions for the existence of an \textit{exact} $M$-dimensional matrix-product stationary state (MPSS)(\S\ref{sec:method-how-to-find-the-condition}),

 \item a systematic way by which the 
exact
$M(\ge N)$-dimensional representation of the the matrix product form(MPF) can
       be constructed from the stationary states of the small $L$ 
if the necessary condition obtained in the abovementioned step 1 is also a sufficient condition for
the existence of an $M$-dimensional MPSS
(\S\ref{sec:method-making-MPF}), 

 \item a systematic way to check the validity of the obtained 
MPSS
for an arbitrary system size in the case
       $N=M$(\S\ref{sec:method-how-to-make-Ec-Dc}). 

\end{enumerate}

After giving the general ideas, we have presented three examples to which our methods can be applied (\S\ref{sec:ex}):
\begin{enumerate}

 \item the asymmetric exclusion process(ASEP)
       \cite{review-of-matrix-product-solution,EsslerRittenberg,MallickSandow1997,Sasamoto-Polynomial-approach-1,Sasamoto-Polynomial-approach-2,B-Polynomial-approach,SS,USW}(\S\ref{sec:example-ASEP}) 

 \item the model in \cite{J2}(\S\ref{sec:J-model-031127-0852})

 \item a hybrid of 1. and 2.(\S\ref{seq:example-hybrid})
\end{enumerate}
For the first two models, we have reproduced the known finite dimensional MPSSs.  For the hybrid model, we have found a
new MPSS by our method.  This clearly shows a potential power of our method to find exact MPSSs for a large class of
one-dimensional stochastic models.

Although we have treated only \textit{exact} MPFs in this paper, we can also show the way how to construct
\textit{numerically} exact MPF for more generic cases where 
calculations are so tedious that the computer algebra system 
(for example, Maple) is not useful 
(this method will be published elsewhere).  It should also be
noted that we have treated \textit{uniform} MPFs in this paper, although \textit{non-uniform} MPF can be made
numerically, for example, by the DMRG\cite{DMR}.

Finally we would like to
mention
possible extensions of our methods.
First step is to remove a restriction($N=M$) on applicability of our methods(\S\ref{sec:method}).
This step helps to develop
 the applicability of our methods
 to various kinds of models(e.g. models with periodic boundary
 conditions; quantum spin chains) and also
to
 develop 
 the numerical renormalization(for example, the DMRG\cite{DMR}) to 
 stochastic models.\cite{Hieida,KP,application-of-DMRG1,application-of-DMRG2,application-of-DMRG3,application-of-DMRG13,application-of-DMRG12,application-of-DMRG4}

\section{Acknowledgement}

Y.H. thanks 
Y. Akutsu, T. Nishino, K. Okunishi and N. Maeshima
for continuous arguments on the matrix product states in the DMRG.  This work was supported by Grant-in-Aid for Young Scientists (B)(14740250).

\vspace{2cm}

\noindent
\textit{Note added.} After the completion of writing the manuscript,
we noticed that our results for the hybrid model, i.e.,
the condition
 (\ref{eq:031213-1554})
and the two-dimensional set of matrices, have already been
derived in \cite{NJP} and \cite{J3} respectively.
They were, however, new when they were discovered and hence
we did not change the original presentation of this article.


\end{document}